\documentclass{article}
\usepackage[T1]{fontenc} 
\usepackage[utf8]{inputenc} 
\usepackage{ismir,amsmath,cite,url}
\usepackage{graphicx}

\usepackage{color}

\def\lbd{}	

\usepackage{lineno}

\title{MidiTok Visualizer: a tool for visualization and analysis of tokenized MIDI symbolic music}

\multauthor
{Michał Wiszenko \hspace{1cm} Kacper Stefański \hspace{1cm} Piotr Malesa} { \bfseries{Łukasz Pokorzyński \hspace{1cm} Mateusz Modrzejewski}\\
Institute of Computer Science, Warsaw University of Technology, Poland\\
{\tt\small michal.wiszenko.stud@pw.edu.pl, kacper.stefanski.stud@pw.edu.pl, piotr.malesa.stud@pw.edu.pl}\\
{\tt\small lukasz.pokorzynski.stud@pw.edu.pl, mateusz.modrzejewski@pw.edu.pl}
}

\def\authorname{M. Wiszenko, K. Stefański, P. Malesa, Ł. Pokorzyński, and M. Modrzejewski}

\begin{document}

\maketitle
\begin{abstract}

Symbolic music research plays a crucial role in music-related machine learning, but MIDI data can be complex for those without musical expertise. To address this issue, we present MidiTok Visualizer, a web application designed to facilitate the exploration and visualization of various MIDI tokenization methods from the \texttt{MidiTok} Python package. MidiTok Visualizer offers numerous customizable parameters, enabling users to upload MIDI files to visualize tokenized data alongside an interactive piano roll.

\end{abstract}
\section{Introduction}\label{sec:introduction}

Symbolic music research plays a key role in advancing machine learning applications in music. However, the inherent structure of MIDI data often presents challenges to AI researchers without musical training or experience in music production. Unlike other data types, such as text, which are more intuitive to tokenize and process in modern models like transformers, MIDI requires a deeper understanding of musical theory and performance conventions. This gap in familiarity can hinder the development of more effective AI models in the field of music generation and analysis. Moreover, various methods for tokenizing MIDI have been proposed, but systematic comparisons of their effectiveness have only recently begun to receive attention in research \cite{fradet2023impact}. 

\texttt{MidiTok} \cite{fradet2023miditok} has been a significant advancement in addressing these issues, as it consolidates multiple MIDI tokenization methods and presents them in a format compatible with transformer-based models. However, understanding these tokenization methods remains challenging for those with limited knowledge of MIDI. To make this more accessible, we introduce MidiTok Visualizer, a tool designed for visualizing and exploring tokenized MIDI files, aimed at both lowering the barrier to entry and providing utility for experienced researchers alike.

\section{Software Overview}
\subsection{Key functionality}
MidiTok Visualizer is a web application designed for visualizing and analyzing MIDI file tokenization techniques from the \texttt{MidiTok} Python package. The key capabilities of the tool are as follows:

\begin{itemize}
    \item Allows users to upload a MIDI file and view a graphical representation of the tokens generated by \texttt{MidiTok}.
    \item Provides an intuitive interface for exploring the structure and content of MIDI data.
    \item Offers a user-friendly interface for configuring parameters related to the underlying \texttt{MidiTok} methods.
    \item Enables users to experiment with different tokenization settings.
    \item Helps users gain deeper insights into the impact of specific parameters on the tokenization outcome.
\end{itemize}

A central feature of the application is a piano roll designed to visualize the notes of the uploaded MIDI file alongside the stream of tokens. It allows highlighting specific tokens to view their corresponding notes within the piano roll and vice versa. Key musical metrics, including key signatures, time signatures, tempo, and pitch range, are displayed above the piano roll. 

MidiTok Visualizer also supports tabs corresponding to individual tracks and includes a built-in player that allows users to listen to the MIDI file.



\subsection{Supported tokenizers}

MidiTok Visualizer currently supports the following tokenizers:
\begin{itemize}
    \item CPWord \cite{cpword2021}
    \item MIDI-Like \cite{oore_midilike_2018}
    \item Octuple \cite{zeng2021musicbert}
    \item REMI \cite{huang_remi_2020}
    \item Structured \cite{pia2021hadjeres}
    \item TSD \cite{fradet-etal-2023-byte}
\end{itemize}
along with dedicated, flexible configurations for each one of them. For a full parameter list for each tokenizer, please refer to the documentation of the original \texttt{MidiTok} \cite{fradet2023miditok}.

\subsection{Design choices}
MidiTok Visualizer has a modular structure and uses FastAPI for the back-end and React for the front-end.

We use \texttt{Pydantic} for data validation and modeling and \texttt{MusPy} \cite{dong2020muspy} for processing MIDI files and extracting key musical information, such as tempo and time signatures. Installing dependencies and running the back-end is simplified using the Poetry tool. We use \texttt{Pytest} for unit testing. 

The application is also containerized with Docker and can be run with a single \texttt{docker compose} command.

\begin{figure}[h]
\centering
\fbox{\includegraphics[width=0.45\textwidth]{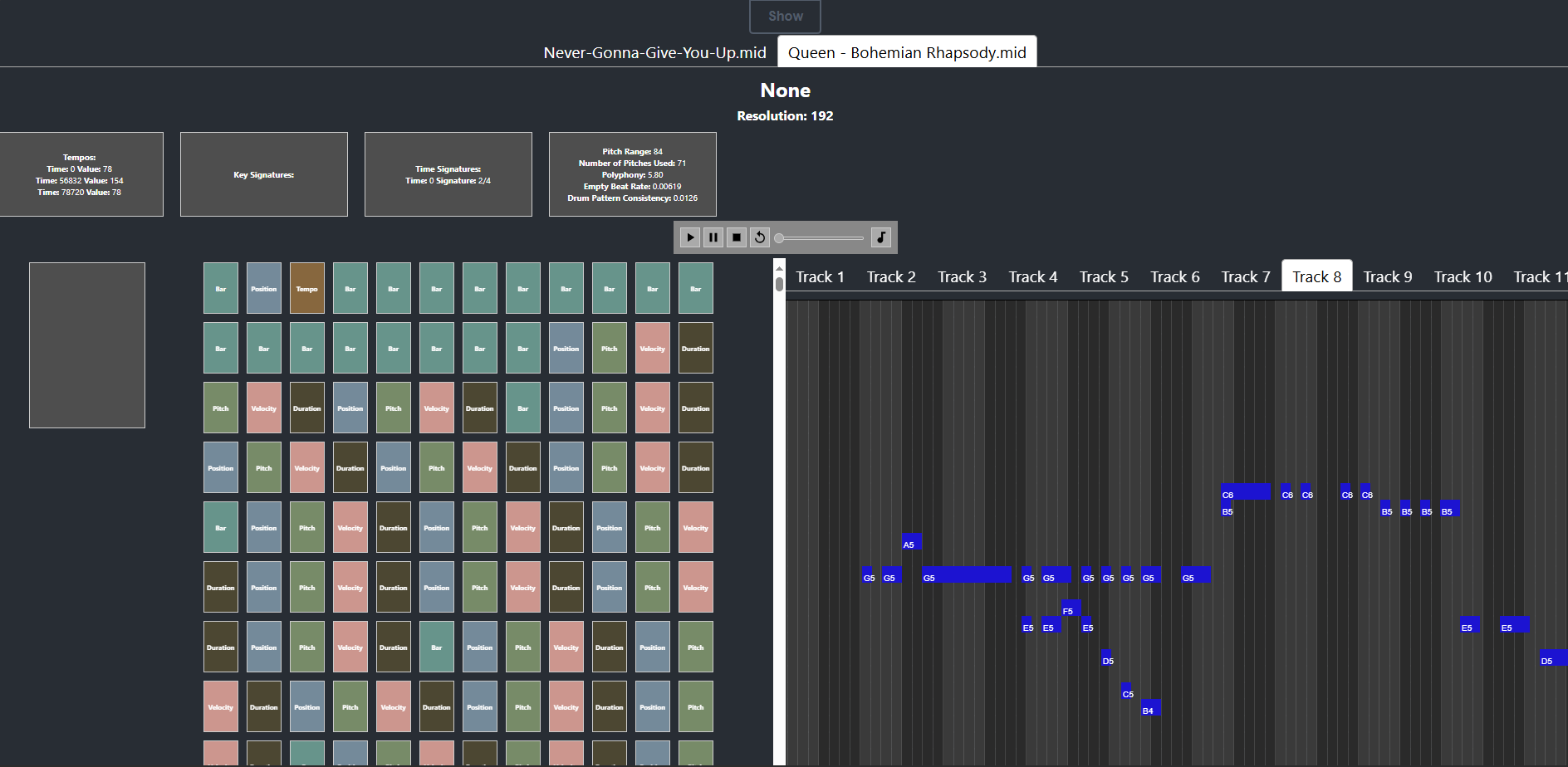}}
\caption{MidiTok Visualizer interface with uploaded file}
\end{figure}


\section{License}

MidiTok Visualizer is released under the GPL license, ensuring it is free for the community to use, modify, and distribute. Contributions from developers, researchers, and enthusiasts are highly encouraged. Whether it's adding new features, fixing existing bugs, or enhancing current capabilities, we welcome collaboration through our public GitHub repository\footnote{https://github.com/justleon/wimu-miditokvisualizer}.

\section{Conclusion and further work}

We introduce MidiTok Visualizer, a functional tool for exploring and visualizing MIDI tokenization methods in a web interface. Its current capabilities allow users to interact with MIDI files and token streams, offering practical utility for both novice and experienced researchers. The tool serves as a foundational platform for further development in the symbolic music research domain, with the goal of increasing accessibility to MIDI data analysis.

Future work will focus on adding support for more tokenizers, handling more complex tokenization scenarios, as well as introducing additional visualization and editing features. 





\bibliography{ISMIRtemplate}

%
%
%
%
%

\end{document}